# Neutron total cross section measurements of gold and tantalum at the nELBE photoneutron source

Roland Hannaske*, Zoltan Elekes, Roland Beyer, Arnd Junghans[1], Daniel Bemmerer, Evert Birgersson[2], Anna Ferrari, Eckart Grosse*, Mathias Kempe*, Toni Kögler*, Michele Marta[3], Ralph Massarczyk*, Andrija Matic[4], Georg Schramm*, Ronald Schwengner, and Andreas Wagner

Helmholtz-Zentrum Dresden-Rossendorf
Bautzner Landstr. 400, 01328 Dresden, Germany

## Abstract

Neutron total cross sections of $^{197}$Au and $^{nat}$Ta have been measured at the nELBE photoneutron source in the energy range from 0.1 – 10 MeV with a statistical uncertainty of up to 2 % and a total systematic uncertainty of 1 %. This facility is optimized for the fast neutron energy range and combines an excellent time structure of the neutron pulses (electron bunch width 5 ps) with a short flight path of 7 m. Because of the low instantaneous neutron flux transmission measurements of neutron total cross sections are possible, that exhibit very different beam and background conditions than found at other neutron sources.

## 1 Introduction

Experimental neutron total cross sections as a function of neutron energy are a fundamental data set for the evaluation of nuclear data libraries. With increasing neutron energy the compound nucleus resonances cannot be resolved anymore and will start to overlap. In the energy range of fast neutrons, which is especially important for innovative nuclear applications, like accelerator driven systems for the transmutation of nuclear waste, the neutron total cross section can be described by optical model calculations (e.g. [1, 2]) where the range below 5 MeV shows a large sensitivity on the optical model parameters.

The neutron total cross section of $^{197}$Au in the energy range from 5 – 200 keV is an item in the OECD NEA Nuclear Data High Priority Request list as $^{197}$Au(n,$\gamma$) is an activation standard in dosimetric applications [3]. Precise total cross section data with a targeted uncertainty < 5 % will have a direct impact on future evaluations of neutron induced reactions on Au. Also an overlapping measurement from 200 keV to 2.5 MeV is of interest to check consistency.

---

[1] Corresponding author: Tel. +49 351 260 3589; email: a.junghans@hzdr.de
* also at Technische Universität Dresden, Dresden, Germany
[2] Present address: AREVA NP GmbH, 91052 Erlangen, Germany
[3] Present address: GSI Helmholtzzentrum für Schwerionenforschung GmbH,
Planckstr. 1, 64291 Darmstadt, Germany
[4] Present address: IBA Particle Therapy, 45157 Essen, Germany

Tantalum is a non-corrosive metal of importance as a structural material in many nuclear and high-temperature applications, e.g. it is also a component in Reduced Activation Ferritic / Martensitic steels [4]. The fast neutron cross section of tantalum has been evaluated recently [5]. In that work, a careful measurement of the neutron total cross section from several tens of keV to several MeV with an accuracy goal of ≈ 1 % has been recommended.

A comprehensive set of very precise high energy neutron total cross sections up to several hundred MeV neutron energy has previously been measured at the Weapons Neutron Research (WNR) spallation neutron source of the Los Alamos National Laboratory (LANL) [6], [7]. Due to experimental constraints these data start at 5 MeV neutron energy, but are still valuable to compare to. At lower energies precise measurements exist using neutrons from the $^7$Li(p,n)$^7$Be reaction at the Fast Neutron Generator of Argonne National Laboratory (ANL) [8]. These data are complemented by the work reported here using a neutron source with very different beam and background properties.

The neutron total cross sections for tantalum of natural isotopic composition (99.95 % purity) and $^{197}$Au (99.99 % purity) were determined by the transmission technique at the photoneutron source nELBE [9] [10] [11] at Helmholtz-Zentrum Dresden-Rossendorf, Germany. This is the world's only neutron time-of-flight facility driven by a superconducting electron accelerator [12] with its superior time structure definition. The very short ( 5 ps) electron bunches allow us to use a short flight path (7.175 m) with a good time resolution and maximize the available neutron intensity with a high repetition rate in continuous-wave (cw) operation (101.5625 kHz micropulse repetition rate). This rate is two to three orders of magnitude higher than the pulsed operation at normal-conducting accelerators. A fast plastic scintillator with low detection threshold [13] was used for the time-of-flight measurements. The neutron spectrum of this facility is characterized in a separate publication [10]. The energy range extends from ≈ 10 keV to 10 MeV, which essentially covers the fission neutron spectrum.

First we shortly describe the nELBE neutron time-of-flight facility and the setup for transmission measurements and then we present the data analysis and discuss the results.

## 2    The nELBE time-of-flight facility

At Helmholtz-Zentrum Dresden Rossendorf the world's only compact photoneutron source at a superconducting electron accelerator dedicated to measurements in the fast neutron range has been built. A compact liquid lead circuit is used as a neutron-producing target. Through this technology the neutron beam intensity is not limited by the heat dissipation inside the target. The technical design including thermo-mechanical parameters of the liquid lead circuit and the beam dump is discussed in Ref. [11]. The electron beam is accelerated to 30 MeV in cw-mode by superconducting cavities. The maximum average beam current at a micropulse rate of 13 MHz is 1 mA. The neutron source strength at the nominal beam current has been calculated with the Monte Carlo N-Particle Transport Code MCNP-4C3 to be $10^{13}$ neutrons/s [9]. The accelerator produces high brilliance beams with variable micropulse repetition rates and duty cycles. The bunch duration is about 5 ps, so that the time-of-flight resolution is not degraded and short flight paths can be used with a high-resolution detection system.

Figure 1 shows the floor plan of the neutron time-of-flight facility.  The electron beam passes through a beryllium window mounted on a stainless-steel vacuum chamber and hits the radiator, consisting of a molybdenum channel confining the liquid lead. The channel has a rhombic cross section with 11 mm side length. The electrons generate bremsstrahlung photons which release neutrons in secondary (γ,n) reactions on lead. These leave the radiator almost isotropically, whereas the angular distributions of electrons and photons are strongly forward-peaked. The collimator axis is located at 95° with respect to the electron beam direction. The collimator and the neutron beam properties at the experimental area have been optimized using MCNP-4C3 in order to maintain the correlation of time-of-flight and neutron energy [9]. The collimator of 2.6 m length contains three replaceable elements of lead and borated polyethylene that are mounted inside a precision steel tube [9].

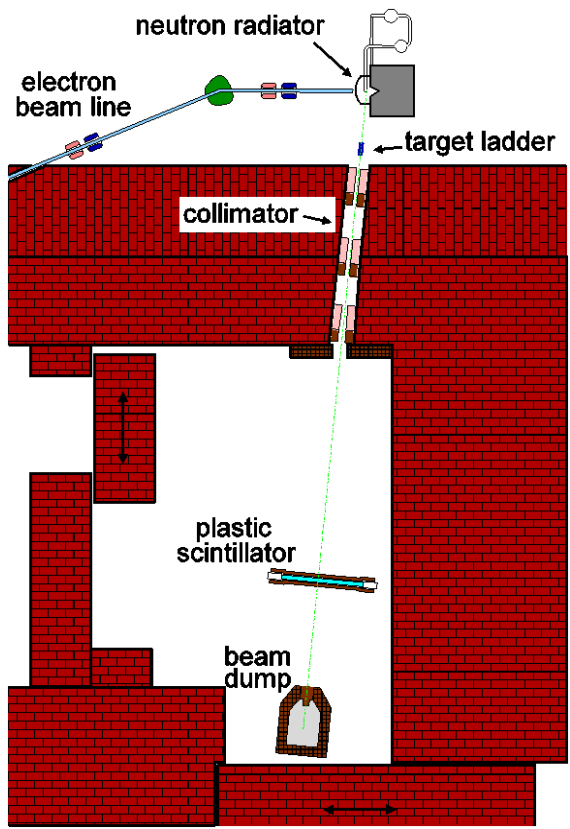

Figure 1: Floor plan of the cave for the neutron transmission experiment at the ELBE accelerator. The neutron radiator consists of a Mo tube with rhombic cross section through which liquid lead is flowing. The target ladder is located approx. 1 m from the neutron radiator in front of the collimator tube which has a length of 2.6 m. The total flight path to the neutron detector (plastic scintillator) is 7.175 m. A beam dump behind the plastic scintillator absorbs the bremsstrahlung and neutrons.

## 3  Transmission experiment

The neutron total cross sections were determined in a transmission experiment. The target samples together with bremsstrahlung absorbers were mounted in a pneumatically driven computer-controlled target ladder directly in front of the collimator entrance. The conical neutron beam collimator has an entrance aperture diameter of 20 mm increasing to 30 mm at the exit. In this geometry small diameter samples were used with a neutron transmission factor of about 0.5. The target samples were periodically moved in and out of the beam to compensate for possible long term drifts in the neutron beam intensity. The data taking time per cycle for the empty sample (3 cm thick Pb bremsstrahlung absorber only) was 600 s, for the Au and Ta samples it was 900 s. The total measurement time was about 78 hours. The order of the cycle was empty-Au-Ta-Pb in the first half of the experiment whereas it was empty-Au-empty-Ta-empty-Pb in the second half with 300 s duration for empty, to increase the frequency of empty target measurements. Each sample was combined with a 3 cm thick Pb absorber to reduce the bremsstrahlung count rate. All Pb absorbers and the Pb sample were made from a technical lead alloy (PbSb4). The data from the Pb sample have been used to determine the energy resolution of the time-of-flight measurement, as shown in section 4.6 .

The transmitted neutrons were detected using a plastic scintillator (Eljen EJ-200, 1000 mm x 42 mm x 11 mm) that was read out on both ends using high-gain Hamamatsu R2059-01 photomultiplier tubes (PMT). The scintillator is surrounded by a 1 cm thick lead shield to reduce the background count rate. The detection threshold for recoil protons in this detector is at about 10 keV [13]. The overlap neutron energy for the given micropulse repetition rate and flight path is 2.8 keV. This is below the detection threshold of the plastic scintillator used in the transmission measurement. The electron beam intensity was reduced to the sub µA range to have a detector count rate of ≈ 10 kHz (empty sample beam). This corresponds to a neutron count rate of ≈250 n/s. On average, only every tenth accelerator bunch is registered by the scintillator.

The time of flight of the transmitted neutrons was measured in list mode with the Multi-Branch-System (MBS[5]) real-time data acquisition developed at GSI, Darmstadt. This setup is optimized to control several VME bus crates with several front-end processors using a real-time operating system. The PMT output signals were fed into a CAEN V874B 4 Channel $BaF_2$ -Calorimeter Read-Out Unit housing charge to digital converter (QDC) and constant fraction discriminator (CFD) sections. An internal constant veto time of ≈2.7 µs in this module helps to suppress the rate of afterpulses that may occur in high-gain PMTs. The CFD output signals were fed into a SIS 3820 scaler module to determine the detector count rate and into the multi-hit multi-event time-to-digital converter (TDC) CAEN V1190A to determine the time information with a dispersion of 97.6 ps/channel. The accelerator radiofrequency (rf)-signal serves as reference for the time-of-flight determination. A CAEN V1495 FPGA module was used to produce the logical AND of the CFD signals from both PMTs to trigger the data acquisition (DAQ). The coincident signals from the TDC that triggered the data acquisition were analyzed for this transmission measurement. The time sum signal of both PMTs is used to measure the time of flight, while a software condition on the time difference signal was used to select events that occurred in the central beam spot region of the scintillator bar. The absolute scale of time of flight is determined using the bremsstrahlung peak and the known flight path. The flight path has been verified by resonant structures that appear in the neutron spectrum [10]: Several absorption minima appear due to resonances with strong elastic neutron scattering in $^{208}$Pb. Emission peaks appear in the neutron spectrum from nuclear levels just above the neutron separation energy mainly in $^{208}$Pb that can be excited via (γ,n) or (e,e'n) reactions.

The transmission T is given by the relation

$$T = \frac{R_{in}}{R_{out}} = \exp(-nl\sigma_{tot}) \tag{1}$$

$$\sigma_{tot} = -\frac{1}{nl}\ln\left(\frac{R_{in}}{R_{out}}\right) = -\frac{1}{nl}\ln(T) \tag{2}$$

---

[5] http://daq.gsi.de

where $R_{in}$, $R_{out}$ are the background and dead-time corrected count rates in the detector with the sample in and out of the beam, respectively. The areal density $nl$ is given by the product of the number density of atoms and the thickness and $\sigma_{tot}$ is the neutron total cross section. The Au and Ta samples used are characterized in Table 1. The areal density $nl$ is known to a relative uncertainty of $6*10^{-3}$.

Table 1: Ta and Au sample characteristics. The samples had cylindrical shape. The density has been calculated from the measured dimensions and mass of the sample to show agreement with the standard density within the relative uncertainty of $6*10^{-3}$ or better. The Ta corresponds to ASTM B365 Grade RO5200, the gold to standard fine gold. The bremsstrahlung absorbers consisted of technical lead alloy (PbSb4) machined to a cylinder (diameter 25.0±0.1 mm, length 30.0±0.1mm).

| Sample | Diameter (mm) | Length (mm) | Mass (g) | Density (g/cm$^3$) | Standard Density (g/cm$^3$) | Purity (weight %) | Areal Density $nl$(atoms/barn) |
|---|---|---|---|---|---|---|---|
| Ta | 25.1±0.1 | 25.5±0.1 | 210.419±0.001 | 16.68±0.09 | 16.65 | 99.95 | 0.1413±0.0006 |
| Au | 26.0±0.1 | 16.0± 0.1 | 163.760±0.001 | 19.28±0.12 | 19.32 | 99.99 | 0.0945±0.0006 |

The spectra of transmitted neutrons measured as a function of time of flight are shown in Figure 2. The flight path from the center of the neutron radiator to the center of the plastic scintillator was determined by geometrical survey to be 717.5 ± 0.2 cm. The bremsstrahlung peak from the neutron radiator has a time of flight of 23.9 ns. The fastest neutrons arrive at about 100 ns after this peak. Measurements with a $^{235}$U fission chamber, which is sensitive down to the thermal region, show that the neutron energy range extends down to about 10 keV [10].

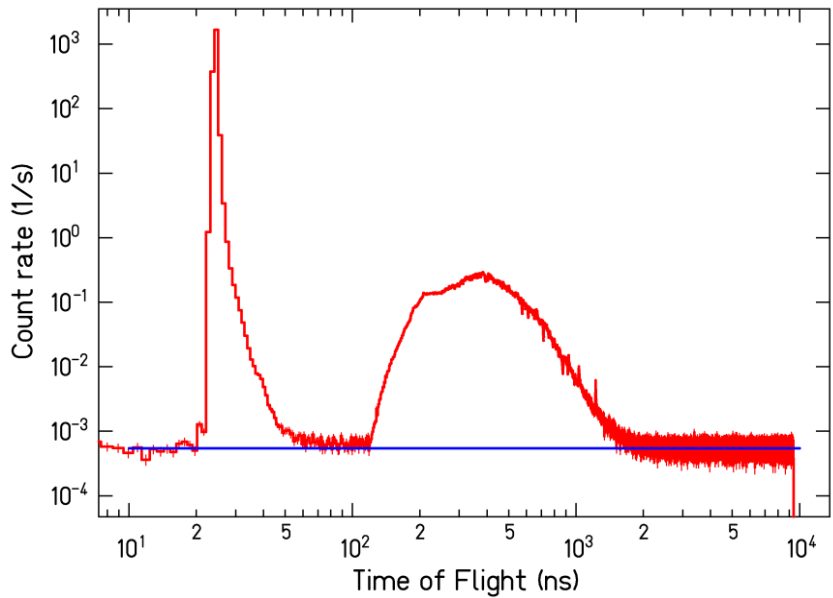

Figure 2: Typical time-of-flight spectrum for the transmission measurement at nELBE. The dead-time corrected count rate is shown as a function of time of flight for the transmission through the Au sample + Pb absorber. A narrow gate was set on the time difference of the two PMTs to select the region, where the transmitted neutron beam passes through the scintillator. The random background rate (blue line) was fitted in the time interval from 8500 ns – 9350 ns. The time resolution of the detection system is characterized by the width of the bremsstrahlung peak and amounts to 0.46 ns.

# 4 Data analysis and experimental uncertainties

To determine the neutron transmission and the total cross section from the measured time-of-flight distribution several corrections have to be done:

1. Correction for a time-of-flight dependent dead time
2. Subtraction of a constant random background in the time-of-flight spectra
3. Neutron beam intensity fluctuations
4. In-scattering of neutrons
5. Resonant self shielding in thick transmission samples

Random background and dead-time corrections are very important. The remaining neutron beam intensity fluctuations were measured in the target cycle and found to have a small influence. In-scattering of neutrons was minimized by using a "good" geometry. Resonant self shielding can be an important correction at low neutron energy, where the total cross section can have strong, separated resonances [8]. Above 100 keV neutron energy this correction is found to be negligible.

## 4.1 Dead-time correction

In time-of-flight measurements at pulsed sources, the dead time correction is a function of the time of flight t, because an event at $t_{start}$ prevents all later signals until $t_{start} + t_{veto}$ from being recorded. Knowing $t_{start}$ and $t_{veto}$, all affected time-of-flight channels *i* can be marked as blocked in a histogram $N_{block}(i)$. This histogram extends from 0 to the accelerator period $T_{acc}$. If a time-of-flight channel beyond the end of the histogram is to be blocked, it is shifted back by an integer number of accelerator periods to fit into the histogram's limits. By this, the overlap of the veto signal into the next accelerator pulses is taken into account. After the measurement the channel-dependent dead-time correction factor can be determined for each time-of-flight channel *i*

$$\alpha(i) = 1 - \frac{N_{block}(i)}{N_{acc}} \qquad (3)$$

in which $N_{acc}$ is the total number of accelerator periods within the measurement. The application of the channel-dependent dead-time correction factor is also described in [10].

In the signal processing and list-mode data acquisition system of this experiment, two independent effects cause dead time, in which detector signals cannot be recorded. The main contribution is the time needed for conversion of the incoming signals to data values and for the transfer of the data to a server. During that time the trigger module is inhibited by a veto signal that is the logical conjunction of the busy signals of the different electronic modules. In contrast to other experiments, in which the dead time has a constant duration [14] or ends at the beginning of the next accelerator pulse [6], $t_{veto}$ is measured event-wise with a CAEN V1495 general purpose VME board using its 40 MHz internal clock, see Figure 3. The time of flight $t_{start}$ is determined from the coincident PMT signals that triggered, see Figure 2. The shape of the resulting dead-time correction factor $\alpha_{DAQ}$ is mainly determined by the 15 µs conversion time of trigger hits from bremsstrahlung photons, which are overlapping into the next accelerator period of about 9.85 µs duration and thus creating a sharp-limited decrease between 0.024 and 5 µs, see Figure 4. In this time-of-flight region, where the neutrons arrive, only very small variations can be seen (55.9 % to 56.0 % in the empty-target measurement). Above 5 µs, where the background is determined, the dead-time correction amounts to 62 %. As the DAQ dead-time correction factor is calculated analytically event by event, there is no related statistical uncertainty. The influence of the time resolution of 25 ns in the measurements of $t_{veto}$ was found to be negligible except for a narrow region around the step at 5 µs.

The channel-dependent DAQ dead time is only caused by events that fulfilled the trigger condition, which is a coincidence of the two PMTs of the plastic scintillation detector. The rate of PMT afterpulses was reduced by the constant veto time of ≈2.7 µs after each PMT signal in the CFD. The drawback of the afterpulse suppression is that also true coincidences occurring during that time are blocked. This is a minor, but still significant contribution (approx. 3 % for the empty target) to the total dead time, see Figure 4. For the calculation of the CFD dead-time correction factors $\alpha_{CFD,1}$ and $\alpha_{CFD,2}$ for each PMT, the method of blocked time-of-flight channels is used again, but now $t_{start}$ and $t_{veto}$ are determined in a different way. The TDC allowed recording all hits 10 µs before and 1.5 µs after each trigger hit. From this information the time-of-flight distribution of non-coincident hits has been determined for each PMT and has to be scaled up to the number of accelerator periods $N_{acc}$ to obtain the $t_{start}$ distribution. The CFD veto length $t_{veto}$ was determined as the minimal difference between subsequent hits and was 2644 ns and 2720 ns, respectively. Due to extrapolations required to obtain the time-of-flight distribution of non-coincident hits over the full range of one accelerator period, there is a contribution of $\alpha_{CFD,1} \cdot \alpha_{CFD,2}$ to the relative uncertainty of the neutron total cross section of about 0.8 %.

Finally, for each time-of-flight channel $i$ the number of truly happened events in the detector $N_{real}(i)$ can be calculated from the number of registered events $N_{live}(i)$

$$N_{real}(i) = \frac{N_{live}(i)}{\alpha_{CFD,1}(i) \cdot \alpha_{CFD,2}(i) \cdot \alpha_{DAQ}(i) \cdot \{1 - \sigma \tanh(\sigma[1 - \alpha_{DAQ}(i)])\}} \tag{4}$$

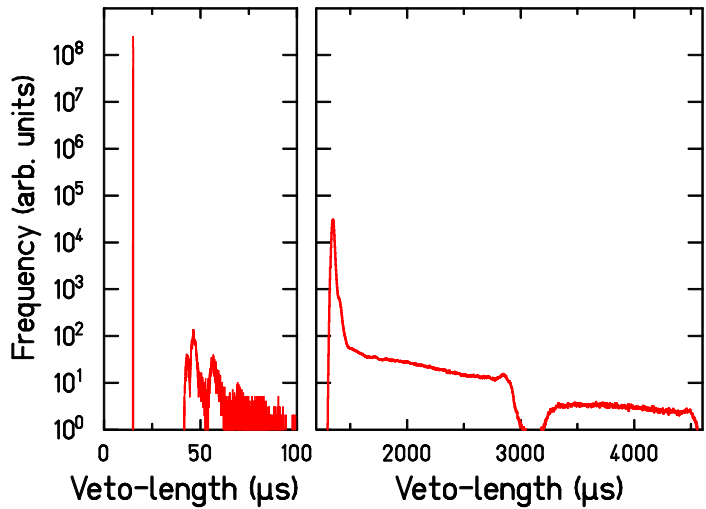

**Figure 3: The per-event dead time has been measured using an FPGA module (CAEN V1495) with an internal 40 MHz clock. The clock was scaled free running and vetoed by the busy signal from the trigger module of the data acquisition, see [10]. The average conversion time per event is approx. 15.1 µs, see left panel. Occasional much longer dead times are due to the data transfer to the server computer, see right panel (The data have been averaged over 25µs.). The average per-event dead time is 59.3 µs.**

The term $1 - \sigma \tanh(\sigma[1 - \alpha_{DAQ}(i)])$ takes beam intensity variations with a relative variance σ² into account [15]. The standard deviation σ is 7.5 % and was determined from the coincident count rate, which was monitored over the full duration of the experiment in intervals of 15 s. The influence of beam intensity variations on the DAQ dead-time correction factor is only 0.3 % (empty) and 0.1 % (Au, Ta). The influence on the CFD dead-time correction factors is negligible.

Transmission measurements typically involve a rather large dead-time correction factor. The correction factor $\alpha_{DAQ}$ described in this section should not suffer from a systematic uncertainty, as the dead time is measured with high time resolution for each event registered in the data acquisition system and the correction factor is applied in a time-of-flight dependent way. It has been shown in this work that the influence of beam intensity fluctuations on the dead-time corrections is very small. They can be treated very well by the method developed by Moore [15]. Experimental tests e.g. with calibration sources, given in [6] [7] [15] have shown that the dead-time correction obtained by this method is correct to a fraction of one percent. In a recent comparison [16] of fast digitizers with a small dead time and a conventional system using a fixed dead time a systematic discrepancy of 2 % was observed for a dead-time correction factor of 0.45 in the conventional system in the extreme case of a saturated resonance in a neutron capture experiment.

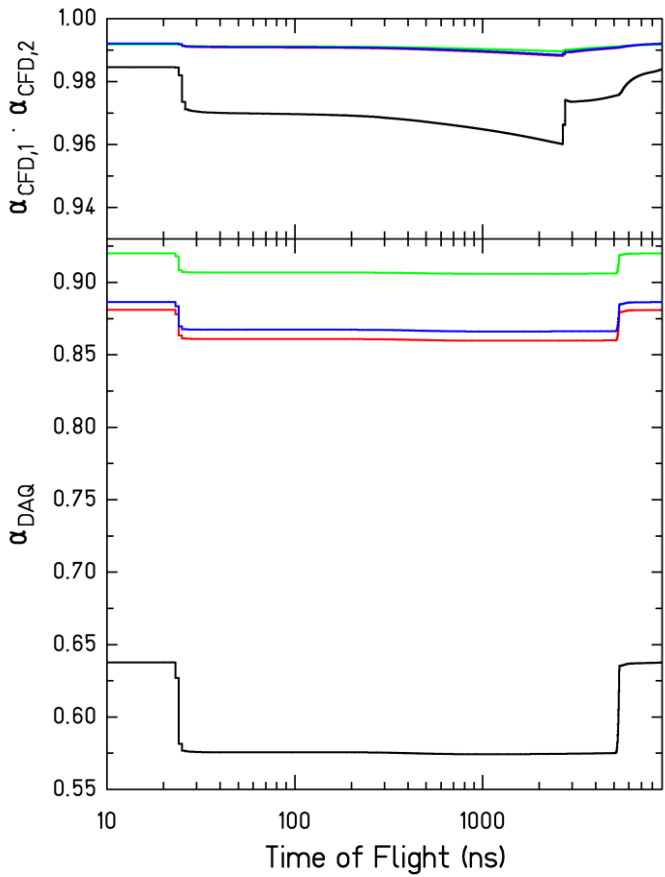

Figure 4: Relative dead-time correction as a function of time of flight in (ns). The lower panel shows the time-of-flight dependent dead-time correction factor $\alpha_{DAQ}$ of the data acquisition system. The dead-time correction for the Ta, Au and Pb samples are shown as green, red and blue lines, the empty target as a black line. The time-of-flight dependence has a characteristic step at about 5 µs from events that were triggered by the bremsstrahlung peak from the penultimate accelerator pulse. The dead-time correction is nearly constant in the neutron time-of-flight range. The upper panel shows the CFD dead-time correction factor $\alpha_{CFD,1} \cdot \alpha_{CFD,2}$ due to the 2.7 µs veto time of the CFD suppressing PMT afterpulses.

## 4.2 Random background correction

The background in a time-of-flight measurement can consist of several components:

- A background independent of the neutron beam intensity and its time structure. This type of background can be determined by a measurement with the electron beam switched off. In this work no background of this type above the usual room background due to ambient radioactivity was found.
- A background related to spurious neutron pulses, e.g. due to small dark currents in the superconducting cavities. This would lead to higher frequency bremsstrahlung peaks that are not observed in the time-of-flight spectra.
- A background related to neutrons scattered in the collimator or in air and detected in the scintillator. This contribution is difficult to determine in the energy range of fast neutrons as a "black" resonance sample, in which the transmission is practically zero around a separated strong resonance, does not exist. The plastic scintillator used has a threshold of about 10 keV neutron energy, so that neutrons that were slowed down below this energy by collisions in air or with the walls of the experimental hall will not be detected. These multiple-scattering slowing down processes tend to take several hundreds of micro-seconds and can finally produce neutron capture gamma rays. With a micropulse repetition rate of 101.5625 kHz a nearly constant background will be produced.

A constant random background was fitted to the dead-time corrected neutron time-of-flight spectra in the time interval above 8500 ns – 9350 ns and subtracted. The result does not depend significantly on the time interval of the fit. The relative systematic uncertainty of the neutron total cross section due to the background determination is 0.2 %. The beam-off random background rate is 87 % of the random background rate for the empty target setting (Pb absorber only). The random background rate in the case of the empty target setting (Pb absorber only) is 11 % higher than in the case of the Au sample + Pb. At neutron energies below 6.4 MeV the subtracted constant background amounts to 1 % or less of the measured count rate (Pb absorber only). Additional measurements using a polyethylene scatterer of 30 cm thickness in the sample position showed that the background is constant in the time-of-flight spectrum at a time of flight higher than 250 ns. At shorter flight times a fraction of fast neutrons was still transmitted.

## 4.3 Neutron beam intensity fluctuations

If the integrated neutron beam intensity is not the same during measurements with sample in and sample out the transmission deduced will not be correct. The ELBE accelerator produces a very stable electron beam, only in a few cases runs with beam instabilities were discarded from the analysis. The plastic scintillator was also used to determine the bremsstrahlung rate and thus to monitor the stability of the electron beam and neutron intensity. The scintillator count rate was measured by a scaler module, which has a dead time much shorter than the accelerator period. As the accelerator pulse duration is shorter than the intrinsic time resolution of the plastic scintillator, the scintillator can give at most one count per accelerator pulse, i.e. multiple bremsstrahlung photons from one accelerator pulse cannot be resolved. From the Poisson distribution the true count rate $n$ (with no detector dead time) can be deduced from the measured count rate $m$, see eq. (4-40) of Ref. [17]:

$$n = f * \ln(\frac{f}{f-m}) \tag{5}$$

with the accelerator frequency $f$ = 101.5625 kHz. The measured average count rate $m$ for "target in" Au and Ta was approximately 2.76 kHz and 1.88 kHz, respectively. For empty target (Pb absorber only) it was 10.1 kHz. The correction for the true count rate $n$ under these conditions is 5 % or smaller. The instantaneous flux of the nELBE photoneutron source is rather small under these conditions. This is remarkably different from the situation at normal conducting accelerators with a very high instantaneous flux. There the measured count rate $m$ will be very close to the accelerator frequency and the true count rate cannot be determined in this way.

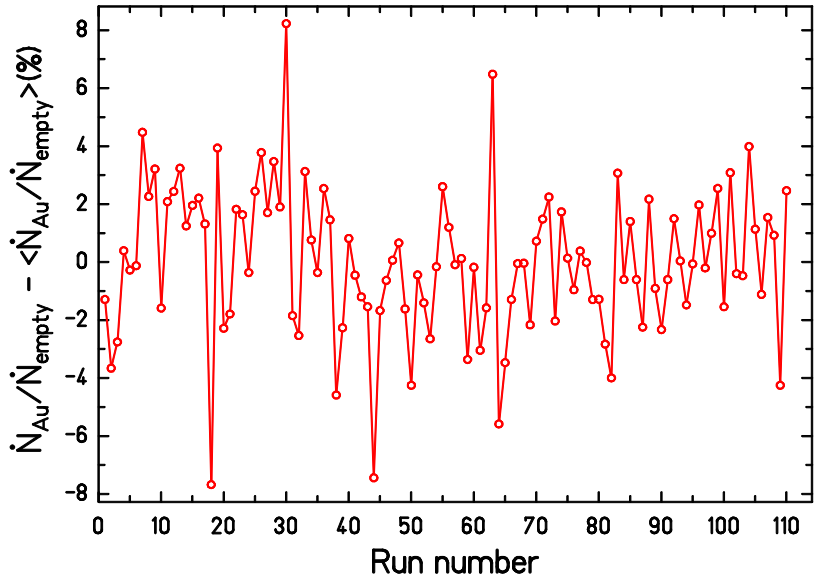

Figure 5: Relative deviation (in percent) of the scaler count rate ratio of each Au sample setting with an adjacent empty setting from the average ratio over the full experiment. The experimental uncertainty of the scaler count rates is smaller than the size of the data symbols.

Figure 5 shows the deviation of the ratio of the scaler count rates of each Au sample and empty setting relative to the average ratio over the full experiment. In the experiment 110 (130) measurements with Au (Ta) were made and 320 with Pb absorber only. The standard deviation of fluctuations from **each** absorber setting to the next is 2.6 %. This value is smaller than the long term fluctuations determined from the scaler rates over the duration of the whole experiment discussed in section 4.1. The transmission spectra and later the neutron total cross sections were determined by summing **all** settings with one absorber to obtain the maximum statistics in the time-of-flight spectrum. To test for short time correlations in the beam intensity transmission spectra were accumulated from 5 consecutive absorber settings with each absorber (Ta+Pb, Au+Pb) and empty (Pb only). The transmission was calculated as the weighted average from these individual transmissions. For the Au(Ta) data the neutron total cross sections determined by these methods agree to 0.5 (0.2) %.

## 4.4 In-scattering correction

A transmission measurement relies on the fact that only neutrons passing through the same thickness of the sample layer without scattering shall be detected. An in-scattering correction could be necessary if neutrons that are multiply scattered by small angles inside the target sample could be detected. A good counting geometry that minimizes this effect is realized by choosing a collimator with a small solid angle. By putting the target samples directly in front of the collimator entrance all neutrons have to pass the sample before they can pass the collimator and reach the detector. The detector also extends a small solid angle relative to the target samples so that the probability to detect scattered neutrons is minimal. The correction for multiple in-scattering of fast neutrons in the Au and Ta samples has been calculated to be negligibly small (< 0.1 %) using cross sections from the JEFF 3.1.1 evaluation with the method of E.M. McMillan and D.C. Sewell, see [18] and references therein.

## 4.5 Resonant self-shielding correction

To achieve a small statistical uncertainty in the transmission measurement rather thick samples with a transmission of about 0.5 have to be used, see Table 1. If the neutron total cross section has a strong resonant structure, self-shielding effects around the resonances can occur. Even in the unresolved resonance range, these corrections can be important on the level of several percent, [8] [19]. The measured transmission is always an average over the experimental energy resolution. Using Eq. (2) one determines an "effective" cross section that is smaller than the real cross section averaged over the experimental resolution. Self-shielding corrections averaged over energy ranges with many resonances can be determined by model calculations, as e.g. included in the SAMMY code [20]. If the real width and strength fluctuations of the neutron resonances are not known, a self-shielding calculation will only yield an ensemble-averaged correction. For the data measured in this experiment (above 100 keV neutron energy), resonant self-shielding corrections were not applied, as the average level spacing in Au and Ta is comparable to the Doppler broadening of the resonances in the room temperature samples. Therefore, the resonances are overlapping and no deviation from exponential attenuation is expected. The Doppler width can be approximated for heavy nuclei and high temperature $T$ [21] [22]

$$\Delta = 2\sqrt{Rk_BT}, \qquad (6)$$

where $R = \frac{m}{M}E_n$ is the recoil energy of the compound nucleus after neutron absorption. The mass of the neutron is denoted by $m$, the mass of the atom by $M$, the neutron kinetic energy by $E_n$ and $k_B$ is the Boltzmann constant. For Au(Ta) the calculated Doppler width is 7.2(7.5) eV at $T$ = 293 K and $E_n$ = 100 keV. The experimental spacing of s-wave resonances in [198]Au (Spins $J$ = 1,2) and [182]Ta (Spins $J$ = 3,4) is 15.7±0.7 eV and 4.17±0.04 eV, respectively [23]. In these nuclei, at 100 keV above the neutron threshold, the average level spacing according to the back shifted Fermi gas model is only less than 0.4 eV smaller than the experimental values, which is still close to the experimental uncertainty. The resonances in the two compound nuclei start to overlap and resonant fluctuations in the cross sections are strongly reduced. In the capture cross section of [197]Au a clustering of resonances has been observed [24] [25]. The average level spacing deduced from the data up to 90 keV neutron energy was 11.2 eV [24]. Using the experimental value of the level spacing from Au implies that above $E_n$ = 240 keV the Doppler width will be equal or larger than the level spacing.

## 4.6 Experimental uncertainties and energy resolution

The statistical uncertainty in the transmission measurement is determined under the assumption of uncorrelated uncertainties including the statistical uncertainty from the sample in/out measurements. Systematic uncertainties include the transmission normalization due to fluctuations in the neutron source intensity, areal density of the target samples and dead-time correction factor. The random background subtraction contributes to both the statistical and systematic uncertainty. The in-scattering and resonant self-absorption are expected to be negligibly small in this measurement. Table 2 shows the statistical uncertainties of the neutron total cross section measurements at selected neutron energies. The statistical uncertainties have been determined from data rebinned to time-of-flight intervals of about 3.9 ns. For the low energy range the data have been rebinned to 97.6 ns time-of-flight intervals. The typical statistical uncertainty amounts to 1-2 % with a systematic (scale factor) uncertainty of about 1 %.

Table 2: Percent uncertainty of the neutron total cross section at selected neutron energies. The upper part shows the statistical uncertainties for selected neutron energies for the Au and Ta measurements. The middle part shows the energy resolution for the neutron energies given above and the bin widths in keV corresponding to the time-of-flight binning used. The values in parentheses at 0.2 MeV are from rebinned data, see also Figure 8. The systematic uncertainties are given in the lower part of the table.

| Statistical uncertainties | | | | |
|---|---|---|---|---|
| $E_n$ (MeV) | 0.2 | 1.0 | 5.0 | 10.0 |
| Ta | 6 % (1.0 %) | 0.9 % | 0.9 % | 2.6 % |
| Au | 8 % (1.4 %) | 1.5 % | 1.2 % | 3.8 % |
| Energy resolution (source + detector) | | | | |
| $\Delta E/E$ (FWHM) | 1.4 % | 1.4 % | 3.5 % | 7.4 % |
| Bin width (keV) | 1.36 (35) | 15.1 | 168 | 465 |
| Systematic uncertainties | | | | |
| Random background subtraction | 0.2 % | | | |
| Transmission normalization | 0.5 % | | | |
| Areal density of the target sample | 0.6 % | | | |
| Dead-time correction factor | 0.8 % | | | |
| Total systematic uncertainty | 1.1 % | | | |

The energy resolution in the fast neutron range, see Figure 5 of Ref. [14], can be described by the time-of-flight to energy correlation of the neutrons that are emitted from the neutron-producing target and pass the collimator and layers of matter in the neutron beam. In a first approximation, the dimensions of the neutron radiator and the time structure of the electron beam that produces the neutrons determine the energy resolution.

To include the effect of neutron scattering, the time-of-flight to energy correlation was simulated with MCNP5 [26] using the full experimental geometry and a realistic photoneutron spectrum, [10] [9]. The number of neutrons detected in a certain time-of-flight interval that lost their original time-of-flight to energy correlation depends on the neutron energy spectrum. Neutrons with higher energy are scattered elastically and inelastically to lower neutron energy where they appear at the same time of flight as unscattered neutrons of a lower energy.

The relative energy resolution of the neutron-producing target at a 700 cm flight path is around $2 \cdot 10^{-3}$ (FWHM) which is mainly dominated by the geometrical extension of the source radiator. Scattering of neutrons in the lead absorber and neutron collimator leads to a moderate increase of the resolution. The energy resolution of the lead shielded plastic scintillator used in this work is larger than expected on the simple estimate taking into account only the detector thickness and time resolution disregarding the scattering of neutrons in the shielding around the detector. A significant contribution comes from neutrons scattered in the shielding material surrounding the neutron detector [10].

The energy resolution has been tested with the 527 keV d-wave resonance in $^{208}$Pb as shown in Figure 6. The effective neutron total cross section has been calculated using Eq. (2) with an experimental transmission taking into account a Gaussian resolution function. An energy resolution of 1.2 % FWHM seems to be realistic. The total width of this resonance is 4.62 keV [23].

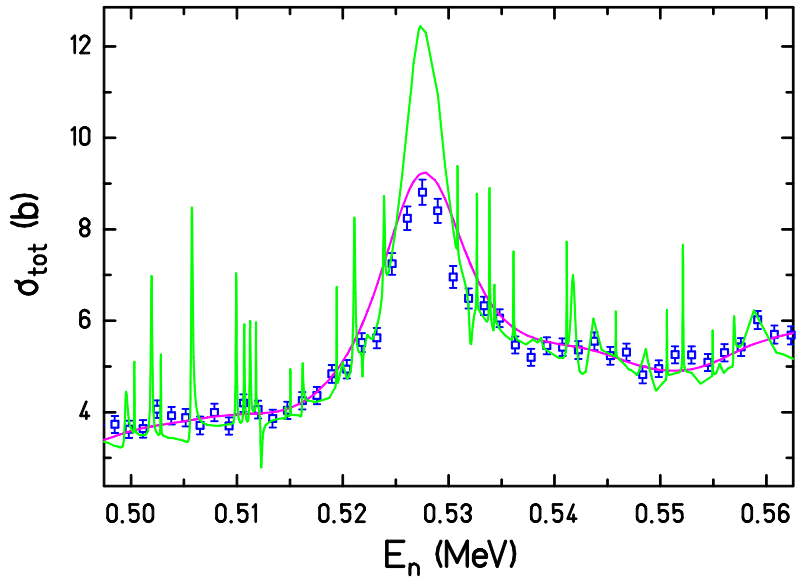

Figure 6: The neutron total cross section measured with a 3 cm thick sample of PbSb4 (technical Pb alloy, blue squares). The resonance at 527 keV neutron energy in $^{208}$Pb can be resolved experimentally. An effective total cross section was calculated from an energy-averaged transmission based on the ENDF/B-VII.1 cross sections (green line) taking into account a Gaussian resolution function with an experimental energy resolution of 1.2 % (FWHM) . The resulting cross section is shown as a purple line.

## 5  Results

The neutron total cross sections of Ta and Au have been measured in the energy range from about 0.1 MeV to 10 MeV. The energy resolution ∆E/E over this energy range increases from 1.4 % to 7.4 % (FWHM). The energy resolution is mostly due to the scattering of neutrons in the lead shield of the plastic scintillator. The resolution is sufficient for average cross sections that can be compared with optical model calculations.

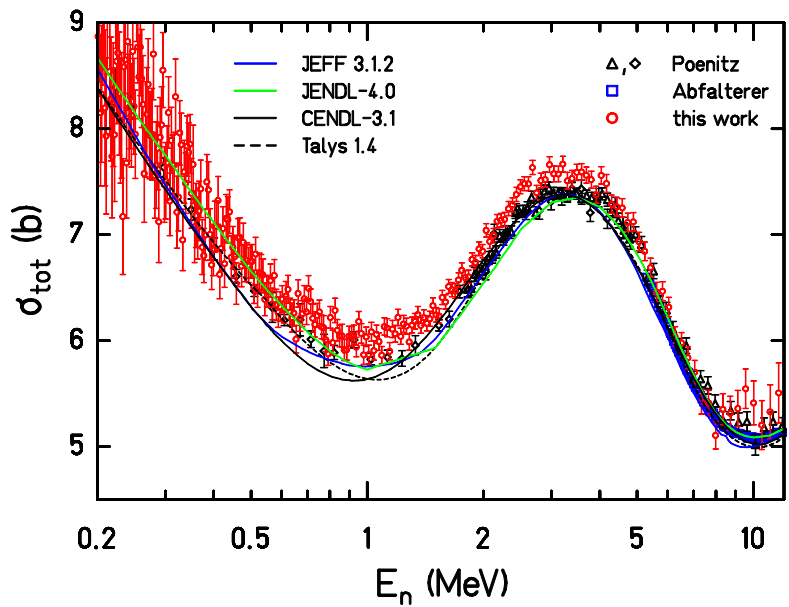

Figure 7: The experimental neutron total cross section of $^{197}$Au as a function of the neutron energy from 0.2 MeV to 10 MeV (this work, red circles). The data from Abfalterer et al. [7] from the LANL WNR spallation neutron source are shown as blue squares. The black symbols denote results from Poenitz et al. [8] [27] from the ANL fast neutron generator. The nELBE data have an equidistant binning in time of 3.9 ns to decrease statistical uncertainties and to increase the readability of the figure. The dashed line shows a result from the Talys code [28]. The JEFF-3.1.2 evaluation is shown by a blue line. The JENDL-4.0 evaluation is shown as a green line and the CENDL-3.1 evaluation as a black line.

In Figure 7 the neutron total cross section of Au is shown in comparison with the data from Poenitz et al. [8] [27] and Abfalterer et al. [7]. The nELBE data are systematically about 2 % higher than the other two experiments. Figure 8 shows a comparison of data in the low energy range below 400 keV. Also here the nELBE data are about 2 % higher than the results from Poenitz et al. The typical statistical uncertainty is smaller than 5 % for Au data at 100 keV and above as was asked for in the High Priority Request List of the Nuclear Energy Agency [3]. Also our measurement extends up to the very accurate data of Abfalterer et al. [7] and thus demonstrates good consistency. The calculated total cross section of the Talys 1.4 reaction code [28] shows good agreement with the data to within 4 %. The Talys code was used with the default options, using the optical model parameters from Koning and Delaroche [1]. Concerning the neutron total cross section of $^{197}$Au all current evaluations are quite similar. The JEFF-3.1.2 and CENDL-3.1 evaluations are about 2-3 % lower than the experimental data. The corresponding ENDF/B-VI.8, ENDF/B-VII.0, and ENDF/B-VII.1 evaluations are identical to JEFF-3.1.2. The JENDL-4.0 evaluation is up to 5 % higher than the evaluations mentioned before.

Figure 9 shows the neutron total cross section of Ta. In the energy range from 0.2 MeV to 10 MeV our data are about 3 % higher than results from Finlay et al. [6] and Poenitz et al. [8] [27] that covered only a part of this energy range. Older data by A.B. Smith [29] are very close in the absolute normalization. A small gap from 0.6 MeV to 1.0 MeV where no high resolution data existed before has been filled. The Talys 1.4 reaction code does not describe the neutron total cross section correctly in the energy range below 2 MeV.

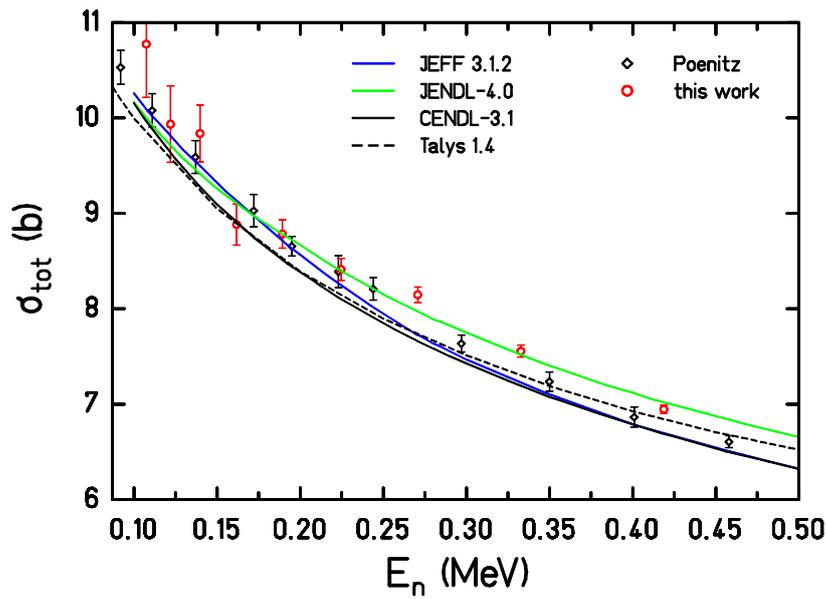

Figure 8: The low energy neutron total cross section of [197]Au. The legend is the same as in the preceding figure. The nELBE data (red circles) tend to be slightly higher than the measurements from Poenitz et al. [8] (black diamonds). The time-of-flight binning has been increased to 97.6 ns to reduce statistical uncertainties and to improve the readability of the figure.

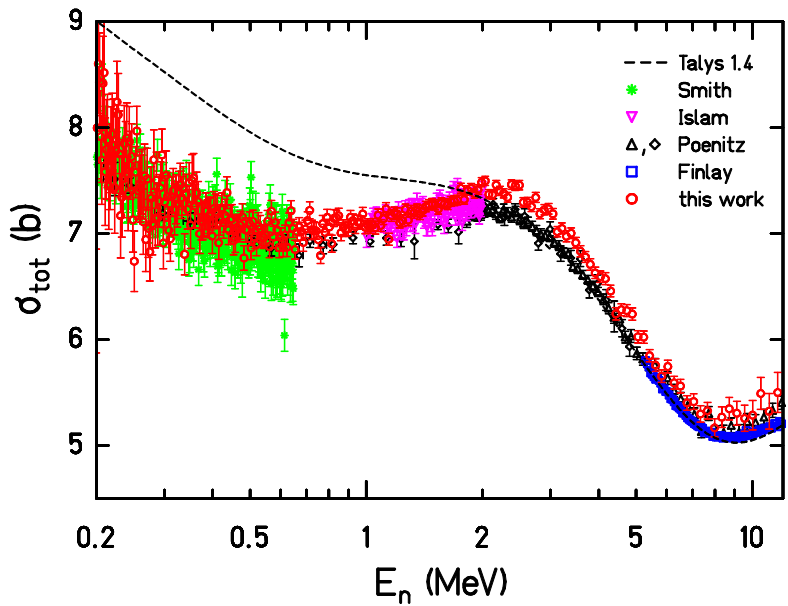

Figure 9: The neutron total cross section for [nat]Ta as a function of the neutron energy from 0.2 MeV to 10 MeV (this work, red circles). The data from Finlay et al. [6] at the LANL WNR spallation neutron source are shown as blue squares. The black symbols denote results from Poenitz et al. [8] [27] from the ANL fast neutron generator. The green and purple symbols denote data from [29] and [30], respectively. The dashed line shows a result from the Talys code [28]. The nELBE data have an equidistant binning in time of 3.9 ns to decrease statistical uncertainties and to increase the readability of the figure.

The neutron total cross section of Ta is compared to different recent nuclear data evaluations in Figure 10.

The JEFF-3.1.2 evaluation is in good agreement with the experimental data, as are the nearly identical curves from RUSFOND 2010 and JENDL-4.0. The ENDF/B-VI.8 and the identical ENDF/B-VII.0 evaluations are below the data. The ENDF/B-VII.1 evaluation is above the data in the energy range below 1 MeV. These discrepancies show again the importance of neutron-total cross section measurements in the fast-energy range covered in this work. This work allows us to base nuclear data evaluations on experimental neutron total cross sections in the energy range below 5 MeV, which is especially sensitive on the optical model parameters used. A careful measurement as recommended in [5] has been done.

The present results with a systematic uncertainty of 1 % might be an indication that the total cross section could be slightly higher than the ones measured before at LANL and ANL. Our detection system had a low detection threshold and good efficiency; however the PMT afterpulsing caused high single count rates that cause additional dead time. To investigate systematic uncertainties in future transmission measurements a data acquisition with much smaller dead-time correction is in preparation. The data measured in this work will be made available through the EXFOR data base. A table of cross sections has been added as electronic supplementary material.

An improved time-of-flight facility is currently under construction in the National Center for High Power Radiation Sources of HZDR Dresden, which includes a 6 m x 6 m x 9 m time-of-flight hall with reduced background from scattering on the walls. This facility will also allow improved measurements of neutron total cross sections to assist future nuclear data evaluations [31].

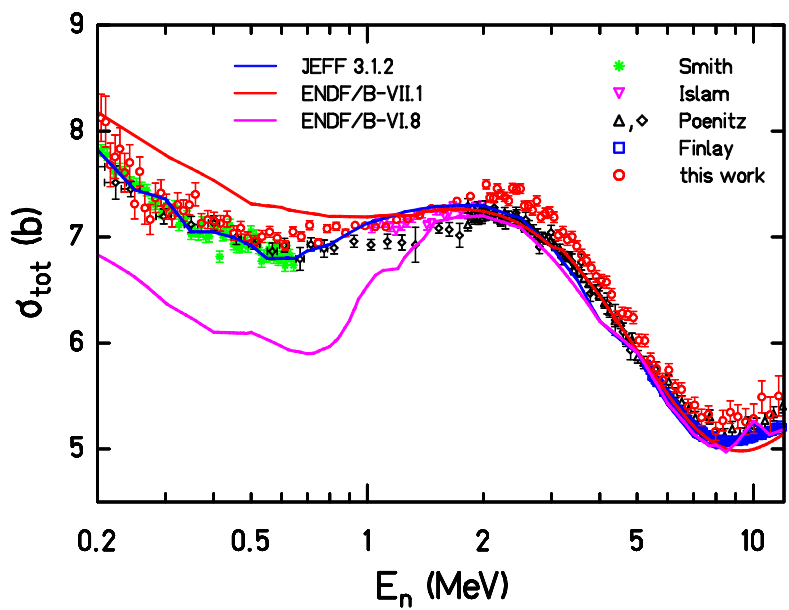

Figure 10: The measured neutron total cross sections for $^{nat}$Ta in comparison with recent nuclear data evaluations. The symbols are the same as in Figure 9. The ENDF/B-VII.1 evaluation is shown by a red curve; the ENDF/B VI.8 evaluation by a purple line. The JEFF-3.1.2 evaluation is shown by a blue line. The experimental data of refs. [29] [30] and this work below 2 MeV have been rebinned to increase readability.

# Acknowledgements

We thank Andreas Hartmann for technical support and preparation of the experiments and the ELBE accelerator crew for providing very stable beam operation. This work is supported by the German Federal Ministry for Education and Science (TRAKULA project, contract number 02NUK13A) and by the European Commission in the projects EFNUDAT (FP6-036434) and ERINDA (FP7-269499).

# Bibliography

[1]   A. Koning and J. Delaroche.*Nucl. Phys. A 713 (2003) 231.*

[2]   P. Pereslavtsev and U. Fischer. *Nucl. Inst. Meth. B 248 (2006) 225.*

[3]   "NEA Nuclear Data High Priority Request List, HPRL," [Online]. Available: http://www.oecd-nea.org/dbdata/hprl/hprlview.pl?ID=428.

[4]   R. L. Klueh, "ELEVATED-TEMPERATURE FERRITIC AND MARTENSITIC STEELS AND THEIR APPLICATION TO FUTURE NUCLEAR REACTORS".*Oak Ridge National Laboratory Report (2004) ORNL/TM-2004/176.*

[5]   A. B. Smith.*Ann. Nucl. Ene. 32 (2005) 1926.*

[6]   R. W. Finlay, W. P. Abfalterer, G. Fink, E. Montei, T. Adami, P. W. Lisowski, G. L. Morgan and R. C. Haight.*Phys. Rev. C 47 (1993) 237.*

[7]   W. P. Abfalterer, F. B. Bateman, F. S. Dietrich, R. W. Finlay, R. C. Haight and G. L. Morgan.*Phys. Rev. C 63 (2001) 044608.*

[8]   W. Poenitz, J. Whalen and A. Smith.*Nucl. Sci. Eng. 78 (1981) 333.*

[9]   J. Klug, E. Altstadt, C. Beckert, R. Beyer, H. Freiesleben, V. Galindo, E. Grosse, A. R. Junghans, D. Legrady, B. Naumann, K. Noack, G. Rusev, K. Schilling, R. Schlenk, S. Schneider, A. Wagner and F.-P. Weiss.*Nucl. Inst. Meth. A 577 (2007) 641.*

[10] R. Beyer, E. Birgersson, Z. Elekes, A. Ferrari, E. Grosse, R. Hannaske, A. Junghans, T. Kögler, R. Massarczyk, A. Matic, R. Nolte, R. Schwengner and A. Wagner.*Nucl. Inst. Meth. A 723 (2013) 151.*

[11] E. Altstadt, C. Beckert, H. Freiesleben, V. Galindo, E. Grosse, A. Junghans, J. Klug, B. Naumann, S. Schneider, R. Schlenk, A. Wagner and F.-P. Weiss.*Ann. Nucl. Ene. 34 (2007) 36.*

[12] F. Gabriel, P. Gippner, E. Grosse, D. Janssen, P. Michel, H. Prade, A. Schamlott, W. Seidel, A. Wolf and R. Wünsch.*Nucl. Inst. Meth. B 161 (2000) 1143.*

[13] R. Beyer, E. Grosse, K. Heidel, J. Hutsch, A. Junghans, J. Klug and D. Legrady.*Nucl. Inst. Meth. A 575 (2007) 449.*

[14] P. Schillebeeckx, B. Becker, Y. Danon, K. Guber, H. Harada, J. Heyse, A. R. Junghans, S. Kopecky, C. Massimi, M. C. Moxon, N. Otuka, I. Sirakov and K. Volev.*Nuclear Data Sheets 113(2012), 3054-3100.*

[15] M. Moore.*Nucl. Inst. Meth. 169 (1980) 245-247.*

[16] L. Mihailescu, A. M. C. Borella and P. Schillebeeckx.*Nuclear Instruments and Methods in Physics Research A 600 (2009) 453–459.*

[17] G. Knoll, Radiation Detection and Measurement, New York: John Wiley & Sons, 1989.

[18] D. Foster and D. Glasgow.*Phys. Rev. C3 (1971) 576.*

[19] H. Derrien, J. Harvey, K. Guber and L. L. N. Leal, *ORNL/TM-2003/291,* 2004.

[20] N. Larson, *ORNL/TM-9179/R6,* July 2003.

[21] W. Lamb Jr..*Phys. Rev. 55 (1939) 190.*

[22] J. Lynn, W. Trela and K. Meggers.*Nucl. Inst. Meth. B 192 (2002) 318.*

[23] S. Mughabghab, Atlas of Neutron Resonances, Amsterdam: Elsevier, 2006.

[24] R. Macklin, J. Halperin and R. Winter.*Phys. Rev. C 11 (1975) 1270.*

[25] C. Lederer, N. Colonna, C. Domingo-Pardo and nToF-Collaboration.*Phys. Rev. C83 (2011) 034608.*

[26] X-5_Monte_Carlo_Team, "MCNP — A General Monte Carlo N-Particle Transport Code, Version 5," *Los Alamos National Lab. Report LA-UR-03-1987,* April 24, 2003 (Revised 10/3/05).

[27] W. Poenitz and J. Whalen, "Neutron Total Cross Section Measurements in the energy region from 47 keV to 20 MeV," Argonne National Laboratory , Argonne, U.S.A., 1983 ANL/NDM-80.

[28] A. Koning, S. Hilaire and M. Duijvestijn, *Proc. Int Conf. Nucl. Data Science and Tech.,April 22-27, 2007 Nice, France, EDP Sciences 2008 p. 211,* .

[29] A. B. Smith, P. T. Guenther and J. F. Whalen.*Phys. Rev. 168 (1968) 1344.*

[30] E. Islam, M. Hussain, N. Ameen, M. Enayetullah and M. Islam.*Nucl. Phys. A209 (1973) 189.*

[31] A. R. Junghans, R. Beyer, Z. Elekes, E. Grosse, R. Hannaske, T. Kögler, R. Massarczyk, R. Schwengner and A. Wagner.*Proc. Int. Conf. Nuclear Data for Science and Technology, New York (2013), to be published in Nuclear Data Sheets (2014).*

Neutron total cross section of $^{197}$Au (time of flight binsize 3.9 ns)

| $E_n$ (MeV) | $\sigma$(barn) | $\Delta\sigma_{stat}$(barn) |
|---|---|---|
| 0.200 | 8.91 | 0.74 |
| 0.201 | 8.87 | 0.71 |
| 0.202 | 9.12 | 0.69 |
| 0.204 | 8.15 | 0.69 |
| 0.205 | 8.33 | 0.64 |
| 0.206 | 9.04 | 0.63 |
| 0.208 | 9.22 | 0.66 |
| 0.209 | 8.20 | 0.74 |
| 0.211 | 7.93 | 0.74 |
| 0.212 | 9.18 | 0.67 |
| 0.214 | 9.68 | 0.63 |
| 0.215 | 8.54 | 0.60 |
| 0.217 | 8.28 | 0.58 |
| 0.218 | 7.74 | 0.58 |
| 0.220 | 8.23 | 0.58 |
| 0.221 | 7.94 | 0.63 |
| 0.223 | 9.37 | 0.61 |
| 0.225 | 8.64 | 0.59 |
| 0.226 | 8.71 | 0.57 |
| 0.228 | 8.33 | 0.58 |
| 0.229 | 8.08 | 0.59 |
| 0.231 | 8.12 | 0.59 |
| 0.233 | 7.17 | 0.55 |
| 0.234 | 8.58 | 0.53 |
| 0.236 | 8.65 | 0.53 |
| 0.238 | 7.68 | 0.52 |
| 0.240 | 8.82 | 0.49 |
| 0.241 | 8.20 | 0.50 |
| 0.243 | 8.30 | 0.49 |
| 0.245 | 8.13 | 0.49 |
| 0.247 | 8.73 | 0.45 |
| 0.249 | 8.21 | 0.45 |
| 0.251 | 7.95 | 0.42 |
| 0.253 | 9.18 | 0.36 |
| 0.254 | 8.77 | 0.34 |
| 0.256 | 8.49 | 0.39 |
| 0.258 | 8.77 | 0.46 |
| 0.260 | 8.47 | 0.46 |
| 0.262 | 8.23 | 0.46 |
| 0.264 | 7.58 | 0.46 |
| 0.267 | 7.99 | 0.45 |
| 0.269 | 8.14 | 0.44 |
| 0.271 | 8.42 | 0.43 |
| 0.273 | 7.94 | 0.41 |
| 0.275 | 7.93 | 0.40 |
| 0.277 | 7.95 | 0.43 |
| 0.279 | 8.70 | 0.43 |
| 0.282 | 7.56 | 0.40 |
| 0.284 | 7.66 | 0.41 |
| 0.286 | 7.82 | 0.39 |
| 0.289 | 7.12 | 0.38 |
| 0.291 | 7.92 | 0.37 |
| 0.293 | 8.04 | 0.36 |
| 0.296 | 8.55 | 0.35 |
| 0.298 | 7.47 | 0.36 |
| 0.301 | 7.51 | 0.36 |
| 0.303 | 7.68 | 0.35 |
| 0.306 | 8.15 | 0.34 |
| 0.308 | 7.32 | 0.33 |
| 0.311 | 7.89 | 0.31 |
| 0.313 | 7.66 | 0.28 |
| 0.316 | 7.66 | 0.31 |
| 0.319 | 8.30 | 0.35 |
| 0.321 | 7.21 | 0.34 |
| 0.324 | 7.35 | 0.32 |
| 0.327 | 7.96 | 0.30 |
| 0.330 | 7.85 | 0.30 |
| 0.333 | 6.95 | 0.29 |
| 0.336 | 7.59 | 0.30 |
| 0.339 | 7.49 | 0.29 |
| 0.342 | 7.84 | 0.28 |
| 0.345 | 7.78 | 0.28 |
| 0.348 | 7.85 | 0.28 |
| 0.351 | 7.39 | 0.33 |
| 0.354 | 7.66 | 0.37 |
| 0.357 | 7.07 | 0.34 |
| 0.360 | 7.18 | 0.30 |
| 0.364 | 7.06 | 0.29 |
| 0.367 | 7.22 | 0.28 |
| 0.370 | 7.43 | 0.27 |
| 0.374 | 6.88 | 0.26 |
| 0.377 | 6.80 | 0.26 |
| 0.381 | 7.00 | 0.24 |
| 0.384 | 7.50 | 0.26 |
| 0.388 | 7.42 | 0.26 |
| 0.392 | 7.00 | 0.24 |
| 0.395 | 6.71 | 0.23 |
| 0.399 | 6.86 | 0.24 |
| 0.403 | 7.08 | 0.24 |
| 0.407 | 7.08 | 0.24 |
| 0.411 | 7.32 | 0.23 |
| 0.415 | 6.62 | 0.23 |
| 0.419 | 6.87 | 0.21 |
| 0.423 | 7.06 | 0.21 |
| 0.427 | 6.73 | 0.22 |
| 0.431 | 7.15 | 0.23 |
| 0.436 | 7.14 | 0.22 |
| 0.440 | 7.22 | 0.20 |
| 0.444 | 6.79 | 0.20 |
| 0.449 | 6.88 | 0.20 |
| 0.453 | 6.79 | 0.20 |
| 0.458 | 6.54 | 0.19 |
| 0.463 | 6.97 | 0.18 |
| 0.468 | 6.98 | 0.18 |
| 0.472 | 6.70 | 0.19 |
| 0.477 | 6.93 | 0.18 |
| 0.482 | 6.73 | 0.17 |
| 0.487 | 6.75 | 0.17 |
| 0.493 | 6.71 | 0.17 |
| 0.498 | 6.75 | 0.16 |
| 0.503 | 6.75 | 0.16 |
| 0.509 | 6.51 | 0.16 |
| 0.514 | 6.82 | 0.16 |
| 0.520 | 6.47 | 0.16 |
| 0.525 | 6.64 | 0.19 |
| 0.531 | 6.63 | 0.18 |
| 0.537 | 6.62 | 0.15 |
| 0.543 | 6.39 | 0.15 |
| 0.549 | 6.47 | 0.15 |
| 0.555 | 6.46 | 0.15 |
| 0.562 | 6.57 | 0.16 |
| 0.568 | 6.35 | 0.15 |
| 0.574 | 6.61 | 0.15 |
| 0.581 | 6.34 | 0.15 |
| 0.588 | 6.27 | 0.14 |
| 0.595 | 6.32 | 0.14 |
| 0.602 | 6.39 | 0.13 |
| 0.609 | 6.29 | 0.12 |
| 0.616 | 6.42 | 0.14 |
| 0.623 | 6.48 | 0.13 |
| 0.631 | 6.27 | 0.13 |
| 0.638 | 6.22 | 0.14 |
| 0.646 | 6.26 | 0.12 |
| 0.654 | 6.01 | 0.13 |
| 0.662 | 6.45 | 0.13 |
| 0.670 | 6.21 | 0.13 |
| 0.679 | 6.25 | 0.12 |
| 0.687 | 6.40 | 0.12 |
| 0.696 | 6.27 | 0.12 |
| 0.705 | 6.49 | 0.12 |
| 0.714 | 6.37 | 0.12 |
| 0.723 | 6.21 | 0.13 |
| 0.732 | 6.15 | 0.12 |
| 0.742 | 6.24 | 0.11 |
| 0.751 | 6.27 | 0.11 |
| 0.761 | 6.13 | 0.11 |
| 0.771 | 6.12 | 0.11 |
| 0.782 | 6.01 | 0.11 |
| 0.792 | 5.88 | 0.11 |
| 0.803 | 6.22 | 0.10 |
| 0.814 | 5.98 | 0.12 |
| 0.825 | 6.29 | 0.12 |
| 0.837 | 6.10 | 0.10 |
| 0.848 | 6.04 | 0.10 |
| 0.860 | 6.05 | 0.11 |
| 0.872 | 5.91 | 0.10 |
| 0.885 | 5.91 | 0.10 |
| 0.897 | 6.04 | 0.09 |
| 0.910 | 6.04 | 0.09 |
| 0.923 | 6.05 | 0.09 |
| 0.937 | 5.94 | 0.09 |
| 0.951 | 6.18 | 0.09 |
| 0.965 | 6.07 | 0.09 |
| 0.979 | 5.87 | 0.09 |
| 0.994 | 6.04 | 0.09 |
| 1.009 | 5.87 | 0.09 |
| 1.025 | 6.08 | 0.09 |
| 1.041 | 5.91 | 0.09 |
| 1.057 | 6.06 | 0.09 |
| 1.073 | 5.99 | 0.08 |
| 1.090 | 5.86 | 0.08 |
| 1.108 | 6.13 | 0.08 |
| 1.125 | 6.12 | 0.08 |
| 1.144 | 5.90 | 0.08 |
| 1.162 | 6.11 | 0.08 |
| 1.181 | 6.07 | 0.08 |
| 1.201 | 6.22 | 0.08 |
| 1.221 | 6.04 | 0.08 |
| 1.242 | 6.16 | 0.08 |
| 1.263 | 6.14 | 0.07 |
| 1.285 | 5.99 | 0.08 |
| 1.307 | 6.15 | 0.07 |
| 1.330 | 6.18 | 0.07 |
| 1.353 | 6.25 | 0.07 |
| 1.377 | 6.14 | 0.07 |
| 1.402 | 6.22 | 0.07 |
| 1.427 | 6.16 | 0.07 |
| 1.453 | 6.21 | 0.07 |
| 1.480 | 6.33 | 0.07 |
| 1.507 | 6.24 | 0.07 |
| 1.536 | 6.23 | 0.07 |
| 1.565 | 6.41 | 0.07 |
| 1.595 | 6.36 | 0.07 |
| 1.626 | 6.51 | 0.07 |
| 1.657 | 6.48 | 0.07 |
| 1.690 | 6.47 | 0.07 |
| 1.724 | 6.57 | 0.07 |
| 1.758 | 6.66 | 0.07 |
| 1.794 | 6.59 | 0.06 |
| 1.831 | 6.60 | 0.06 |
| 1.869 | 6.77 | 0.06 |
| 1.908 | 6.74 | 0.06 |
| 1.948 | 6.87 | 0.07 |
| 1.990 | 6.84 | 0.06 |
| 2.033 | 7.04 | 0.06 |
| 2.077 | 6.94 | 0.06 |
| 2.123 | 7.00 | 0.06 |
| 2.171 | 7.13 | 0.06 |
| 2.220 | 7.13 | 0.06 |
| 2.270 | 7.14 | 0.07 |
| 2.323 | 7.24 | 0.07 |
| 2.377 | 7.28 | 0.07 |
| 2.434 | 7.38 | 0.07 |
| 2.492 | 7.45 | 0.07 |
| 2.552 | 7.48 | 0.07 |
| 2.615 | 7.48 | 0.07 |
| 2.680 | 7.63 | 0.07 |
| 2.748 | 7.39 | 0.07 |
| 2.818 | 7.51 | 0.07 |
| 2.890 | 7.52 | 0.07 |
| 2.966 | 7.66 | 0.07 |
| 3.045 | 7.57 | 0.08 |
| 3.126 | 7.57 | 0.07 |
| 3.212 | 7.67 | 0.08 |
| 3.300 | 7.51 | 0.08 |
| 3.393 | 7.53 | 0.08 |
| 3.489 | 7.59 | 0.08 |
| 3.590 | 7.59 | 0.08 |
| 3.695 | 7.56 | 0.08 |
| 3.804 | 7.52 | 0.08 |
| 3.919 | 7.38 | 0.08 |
| 4.039 | 7.56 | 0.08 |
| 4.164 | 7.48 | 0.08 |
| 4.296 | 7.29 | 0.09 |
| 4.434 | 7.26 | 0.09 |
| 4.578 | 7.25 | 0.09 |
| 4.730 | 7.12 | 0.09 |
| 4.889 | 7.12 | 0.09 |
| 5.057 | 6.93 | 0.09 |
| 5.233 | 6.84 | 0.09 |
| 5.419 | 6.69 | 0.09 |
| 5.615 | 6.46 | 0.09 |
| 5.822 | 6.47 | 0.09 |
| 6.041 | 6.32 | 0.09 |
| 6.272 | 6.04 | 0.09 |
| 6.517 | 5.98 | 0.10 |
| 6.776 | 5.84 | 0.10 |
| 7.051 | 5.70 | 0.11 |
| 7.344 | 5.42 | 0.12 |
| 7.655 | 5.41 | 0.12 |
| 7.986 | 5.11 | 0.13 |
| 8.339 | 5.35 | 0.15 |
| 8.717 | 5.32 | 0.16 |
| 9.121 | 5.36 | 0.17 |
| 9.554 | 5.55 | 0.19 |
| 10.018 | 5.41 | 0.21 |

Neutron total cross section of $^{197}$Au  (time of flight binsize 97.6 ns)

| $E_n$ (MeV) | $\sigma$(barn) | $\Delta\sigma_{stat}$(barn) |
|---|---|---|
| 0.107 | 10.77 | 0.56 |
| 0.122 | 9.94 | 0.40 |
| 0.140 | 9.84 | 0.30 |
| 0.162 | 8.88 | 0.22 |
| 0.189 | 8.78 | 0.15 |
| 0.225 | 8.41 | 0.12 |
| 0.271 | 8.15 | 0.08 |
| 0.333 | 7.56 | 0.06 |
| 0.419 | 6.95 | 0.04 |

Neutron total cross section of $^{nat}$Ta  (time of flight binsize 3.9 ns)

| $E_n$ (MeV) | $\sigma$(barn) | $\Delta\sigma_{stat}$(barn) |
|---|---|---|
| 0.200 | 6.36 | 0.49 |
| 0.201 | 8.00 | 0.48 |
| 0.202 | 8.60 | 0.47 |
| 0.204 | 8.43 | 0.46 |
| 0.205 | 7.91 | 0.43 |
| 0.206 | 7.69 | 0.42 |
| 0.208 | 8.42 | 0.44 |
| 0.209 | 8.52 | 0.51 |
| 0.211 | 7.58 | 0.51 |
| 0.212 | 7.79 | 0.47 |
| 0.214 | 7.88 | 0.44 |
| 0.215 | 7.45 | 0.41 |
| 0.217 | 7.63 | 0.40 |
| 0.218 | 7.79 | 0.41 |
| 0.220 | 7.94 | 0.39 |
| 0.221 | 7.83 | 0.44 |
| 0.223 | 7.43 | 0.43 |
| 0.225 | 7.78 | 0.40 |
| 0.226 | 8.24 | 0.38 |
| 0.228 | 8.12 | 0.39 |
| 0.229 | 7.75 | 0.40 |
| 0.231 | 7.17 | 0.40 |
| 0.233 | 7.34 | 0.38 |
| 0.234 | 7.52 | 0.37 |
| 0.236 | 7.85 | 0.36 |
| 0.238 | 7.70 | 0.36 |
| 0.240 | 7.90 | 0.33 |
| 0.241 | 7.38 | 0.35 |
| 0.243 | 7.87 | 0.33 |
| 0.245 | 7.64 | 0.32 |
| 0.247 | 7.73 | 0.31 |
| 0.249 | 7.13 | 0.30 |
| 0.251 | 6.74 | 0.29 |
| 0.253 | 7.58 | 0.24 |
| 0.254 | 7.97 | 0.23 |
| 0.256 | 7.41 | 0.27 |
| 0.258 | 7.47 | 0.31 |
| 0.260 | 7.42 | 0.31 |
| 0.262 | 7.49 | 0.31 |
| 0.264 | 7.11 | 0.31 |
| 0.267 | 7.54 | 0.30 |
| 0.269 | 6.94 | 0.30 |
| 0.271 | 7.51 | 0.29 |
| 0.273 | 7.05 | 0.27 |
| 0.275 | 6.83 | 0.27 |
| 0.277 | 7.34 | 0.29 |
| 0.279 | 7.09 | 0.29 |
| 0.282 | 7.70 | 0.28 |
| 0.284 | 7.14 | 0.28 |
| 0.286 | 7.50 | 0.27 |
| 0.289 | 7.15 | 0.26 |
| 0.291 | 7.31 | 0.25 |
| 0.293 | 7.43 | 0.25 |
| 0.296 | 7.74 | 0.24 |
| 0.298 | 6.91 | 0.24 |
| 0.301 | 7.30 | 0.25 |
| 0.303 | 7.55 | 0.23 |
| 0.306 | 7.35 | 0.23 |
| 0.308 | 7.03 | 0.22 |
| 0.311 | 6.95 | 0.21 |
| 0.313 | 7.35 | 0.19 |
| 0.316 | 7.40 | 0.21 |
| 0.319 | 7.23 | 0.24 |
| 0.321 | 7.32 | 0.23 |
| 0.324 | 7.28 | 0.21 |
| 0.327 | 7.38 | 0.21 |
| 0.330 | 7.46 | 0.20 |
| 0.333 | 7.34 | 0.20 |
| 0.336 | 7.43 | 0.20 |
| 0.339 | 6.80 | 0.19 |
| 0.342 | 7.37 | 0.19 |
| 0.345 | 7.23 | 0.19 |
| 0.348 | 7.48 | 0.19 |
| 0.351 | 7.08 | 0.22 |
| 0.354 | 7.50 | 0.25 |
| 0.357 | 7.53 | 0.23 |
| 0.360 | 7.26 | 0.21 |
| 0.364 | 7.38 | 0.20 |
| 0.367 | 7.30 | 0.19 |
| 0.370 | 7.00 | 0.19 |
| 0.374 | 7.11 | 0.17 |
| 0.377 | 7.11 | 0.17 |
| 0.381 | 7.14 | 0.16 |
| 0.384 | 7.21 | 0.17 |
| 0.388 | 7.14 | 0.18 |
| 0.392 | 7.17 | 0.17 |
| 0.395 | 7.19 | 0.16 |
| 0.399 | 6.98 | 0.16 |
| 0.403 | 7.24 | 0.16 |
| 0.407 | 7.21 | 0.16 |
| 0.411 | 7.20 | 0.15 |
| 0.415 | 6.91 | 0.16 |
| 0.419 | 7.13 | 0.14 |
| 0.423 | 7.24 | 0.14 |
| 0.427 | 7.10 | 0.15 |
| 0.431 | 7.17 | 0.15 |
| 0.436 | 7.39 | 0.15 |
| 0.440 | 7.04 | 0.14 |
| 0.444 | 7.13 | 0.14 |
| 0.449 | 7.14 | 0.13 |
| 0.453 | 6.93 | 0.13 |
| 0.458 | 6.95 | 0.13 |
| 0.463 | 7.06 | 0.12 |
| 0.468 | 7.04 | 0.12 |
| 0.472 | 7.08 | 0.13 |
| 0.477 | 7.16 | 0.12 |
| 0.482 | 6.77 | 0.12 |
| 0.487 | 7.03 | 0.11 |
| 0.493 | 7.03 | 0.11 |
| 0.498 | 7.10 | 0.11 |
| 0.503 | 6.86 | 0.11 |
| 0.509 | 7.11 | 0.11 |
| 0.514 | 6.97 | 0.11 |
| 0.520 | 6.90 | 0.11 |
| 0.525 | 7.07 | 0.13 |
| 0.531 | 7.05 | 0.12 |
| 0.537 | 6.97 | 0.10 |
| 0.543 | 6.96 | 0.10 |
| 0.549 | 6.98 | 0.10 |
| 0.555 | 7.12 | 0.10 |
| 0.562 | 6.88 | 0.11 |
| 0.568 | 7.21 | 0.10 |
| 0.574 | 6.97 | 0.10 |
| 0.581 | 7.10 | 0.10 |
| 0.588 | 7.03 | 0.10 |
| 0.595 | 6.87 | 0.09 |
| 0.602 | 6.97 | 0.09 |
| 0.609 | 6.87 | 0.09 |
| 0.616 | 6.99 | 0.09 |
| 0.623 | 6.87 | 0.09 |
| 0.631 | 7.10 | 0.09 |
| 0.638 | 7.22 | 0.09 |
| 0.646 | 7.05 | 0.08 |
| 0.654 | 6.85 | 0.09 |
| 0.662 | 7.02 | 0.09 |
| 0.670 | 7.06 | 0.09 |
| 0.679 | 7.02 | 0.08 |
| 0.687 | 6.92 | 0.08 |
| 0.696 | 7.04 | 0.08 |
| 0.705 | 7.12 | 0.08 |
| 0.714 | 7.22 | 0.08 |
| 0.723 | 7.03 | 0.09 |
| 0.732 | 7.03 | 0.08 |
| 0.742 | 7.03 | 0.08 |
| 0.751 | 6.94 | 0.08 |
| 0.761 | 6.79 | 0.08 |
| 0.771 | 7.09 | 0.08 |
| 0.782 | 7.01 | 0.08 |
| 0.792 | 7.15 | 0.07 |
| 0.803 | 7.06 | 0.07 |
| 0.814 | 6.91 | 0.08 |
| 0.825 | 7.14 | 0.08 |
| 0.837 | 7.15 | 0.07 |
| 0.848 | 7.17 | 0.07 |
| 0.860 | 7.01 | 0.07 |
| 0.872 | 7.06 | 0.07 |
| 0.885 | 7.04 | 0.07 |
| 0.897 | 7.04 | 0.06 |
| 0.910 | 7.15 | 0.06 |
| 0.923 | 7.12 | 0.06 |
| 0.937 | 7.06 | 0.06 |
| 0.951 | 7.11 | 0.07 |
| 0.965 | 7.10 | 0.06 |
| 0.979 | 7.18 | 0.06 |
| 0.994 | 7.06 | 0.06 |
| 1.009 | 7.08 | 0.06 |
| 1.025 | 7.06 | 0.06 |
| 1.041 | 7.08 | 0.06 |
| 1.057 | 7.06 | 0.06 |
| 1.073 | 7.17 | 0.06 |
| 1.090 | 7.08 | 0.06 |
| 1.108 | 7.14 | 0.06 |
| 1.125 | 7.23 | 0.06 |
| 1.144 | 6.95 | 0.06 |
| 1.162 | 7.17 | 0.06 |
| 1.181 | 7.20 | 0.05 |
| 1.201 | 7.19 | 0.05 |
| 1.221 | 7.10 | 0.05 |
| 1.242 | 7.12 | 0.06 |
| 1.263 | 7.23 | 0.05 |
| 1.285 | 7.13 | 0.05 |
| 1.307 | 7.21 | 0.05 |
| 1.330 | 7.20 | 0.05 |
| 1.353 | 7.23 | 0.05 |
| 1.377 | 7.18 | 0.05 |
| 1.402 | 7.22 | 0.05 |
| 1.427 | 7.25 | 0.05 |
| 1.453 | 7.27 | 0.05 |
| 1.480 | 7.28 | 0.05 |
| 1.507 | 7.27 | 0.05 |
| 1.536 | 7.23 | 0.05 |
| 1.565 | 7.29 | 0.05 |
| 1.595 | 7.28 | 0.05 |
| 1.626 | 7.27 | 0.05 |
| 1.657 | 7.30 | 0.05 |
| 1.690 | 7.31 | 0.05 |
| 1.724 | 7.32 | 0.05 |
| 1.758 | 7.43 | 0.05 |
| 1.794 | 7.32 | 0.04 |
| 1.831 | 7.41 | 0.04 |
| 1.869 | 7.43 | 0.04 |
| 1.908 | 7.38 | 0.04 |
| 1.948 | 7.41 | 0.04 |
| 1.990 | 7.48 | 0.04 |
| 2.033 | 7.50 | 0.04 |
| 2.077 | 7.37 | 0.04 |
| 2.123 | 7.37 | 0.04 |
| 2.171 | 7.46 | 0.04 |
| 2.220 | 7.40 | 0.04 |
| 2.270 | 7.38 | 0.04 |
| 2.323 | 7.34 | 0.05 |
| 2.377 | 7.46 | 0.04 |
| 2.434 | 7.45 | 0.05 |
| 2.492 | 7.46 | 0.04 |
| 2.552 | 7.29 | 0.04 |
| 2.615 | 7.30 | 0.04 |
| 2.680 | 7.25 | 0.05 |
| 2.748 | 7.34 | 0.05 |
| 2.818 | 7.19 | 0.05 |
| 2.890 | 7.22 | 0.05 |
| 2.966 | 7.23 | 0.05 |
| 3.045 | 7.19 | 0.05 |
| 3.126 | 7.02 | 0.05 |
| 3.212 | 7.05 | 0.05 |
| 3.300 | 6.97 | 0.05 |
| 3.393 | 7.00 | 0.05 |
| 3.489 | 6.91 | 0.05 |
| 3.590 | 6.83 | 0.05 |
| 3.695 | 6.78 | 0.05 |
| 3.804 | 6.71 | 0.05 |
| 3.919 | 6.63 | 0.05 |
| 4.039 | 6.59 | 0.05 |
| 4.164 | 6.57 | 0.06 |
| 4.296 | 6.46 | 0.06 |
| 4.434 | 6.24 | 0.06 |
| 4.578 | 6.28 | 0.06 |
| 4.730 | 6.27 | 0.06 |
| 4.889 | 6.24 | 0.06 |
| 5.057 | 6.03 | 0.06 |
| 5.233 | 6.02 | 0.06 |
| 5.419 | 5.85 | 0.06 |
| 5.615 | 5.76 | 0.06 |
| 5.822 | 5.65 | 0.06 |
| 6.041 | 5.75 | 0.06 |
| 6.272 | 5.59 | 0.06 |
| 6.517 | 5.56 | 0.06 |
| 6.776 | 5.40 | 0.07 |
| 7.051 | 5.42 | 0.07 |

| 7.344 | 5.30 | 0.08 |
| 7.655 | 5.33 | 0.08 |
| 7.986 | 5.17 | 0.09 |
| 8.339 | 5.27 | 0.10 |
| 8.717 | 5.35 | 0.11 |
| 9.121 | 5.31 | 0.11 |
| 9.554 | 5.26 | 0.12 |
| 10.018 | 5.29 | 0.14 |